# Prevention of Traumatic Brain Injury with Liquid Shock Absorption

Hossein Vahid Alizadeh[1], Michael G. Fanton[2], August G. Domel[1],

Gerald Grant[3], David Benjamin Camarillo[1*]

**Abstract**

The discovery that repetitive mild traumatic brain injury (mTBI) can result in chronic traumatic encephalopathy (CTE) in high risk contact sports has led to increased scrutiny of head protective gear. In this work, we asked if it was physically possible to prevent mTBI in American football with helmets alone. Here, we show that modern helmets of several types are unlikely to prevent mTBI from high speed collisions as might be seen in the NFL, but that introducing liquid as an energy absorbing medium can dramatically reduce the forces of impact across a spectrum of impact severities. We hypothesized that a helmet which transmits a nearly constant force during football impacts is sufficient to reduce biomechanical loading in the brain below the threshold of mTBI. To test this hypothesis, we first show that the optimal impact force transmitted to the head, in terms of brain strain, is in fact a constant force profile. Then, to generate a constant force with a helmet, we implement a computational model of a fluid-based shock absorber that adapts passively to any given impact speed. Computer simulation of head impacts with liquid shock absorption indicate that, at the highest impact speed, the average brain tissue strain is reduced by 27.6% ± 9.3 compared to existing helmet padding that is available on the market. These simulations are based on the NFL's helmet test protocol and predict that adding liquid shock absorbers could reduce the number of concussions by at least 75%. Taken together, these results suggest that the majority of mTBI in football could be prevented with more efficient helmet technology.

[1] Bioengineering Department, Stanford University, Stanford, CA 94305. [2] Department of Mechanical Engineering, Stanford University, Stanford, CA 94305. [3] Department of Neurosurgery, Stanford University, Stanford, CA94305.
[*]email: dcamarillo@stanford.edu

# Introduction

Concussion is defined as a clinical syndrome characterized by immediate and transient alteration in brain function, including alteration of mental status and level of consciousness, resulting from mechanical force or trauma. Nearly 3.8 million people in the United States alone each year sustain concussions from sports and other recreational activities, affecting football players, cyclists, and both professional and amateur athletes alike [1], [2]. This type of brain injury is a serious public health concern with more than 40 million mTBIs occurring each year worldwide [3], [4]. mTBI can be caused by a physical trauma which occurs in one of two ways: either due to a focal or a diffuse injury [5], [6], [7]–[12]. Focal trauma is caused by a concentrated force leading to skull fracture or deformation [13]. In contrast, diffuse trauma is caused by an inertial loading within the brain tissue due to an acceleration or deceleration [14], [15], [16]. In the presence of a helmet, focal brain trauma is mitigated since modern helmets more effectively distribute concentrated forces on the head. On the other hand, diffuse trauma is still highly widespread, despite helmet use. Although only a single mTBI may not cause serious long-term health consequences, repeated concussive or even sub-concussive head impacts may lead to neurodegeneration and permanent neuropsychological sequelae [17], [18]. Since children have a larger head-to-body weight ratio, thinner protective skull and immature brain that is still myelinating, they may be more susceptible to head injury [19]. Therefore, repeated mTBI in children is a serious concern as it can lead to permanent harm [20].

American football itself is a stark example of the need for a solution to protect athletes from mTBI, given that each football player is hit in the head approximately 1,000 times over the course of a season [21], [22], [30]. Concussion is highly prevalent in football despite helmet use [23], [24]. Every year nearly 100,000 cases of concussion are reported among approximately 2.5 million helmeted football players in the United States, which includes players in Pop Warner league, high school, college, the National Football League (NFL), and Arena football [25]. Although new helmet technology has been introduced in the past decade aimed at reducing the risk of brain injury, concussions in football continue to rise, as evident by the more than 5-fold increase from 1998 to 2015 in high school boy's football [26], [27]. Concussion statistics confirm that current helmets are not effective enough in protecting players from mTBI and demonstrates the critical need to develop helmets with a higher safety performance to mitigate concussion risk.

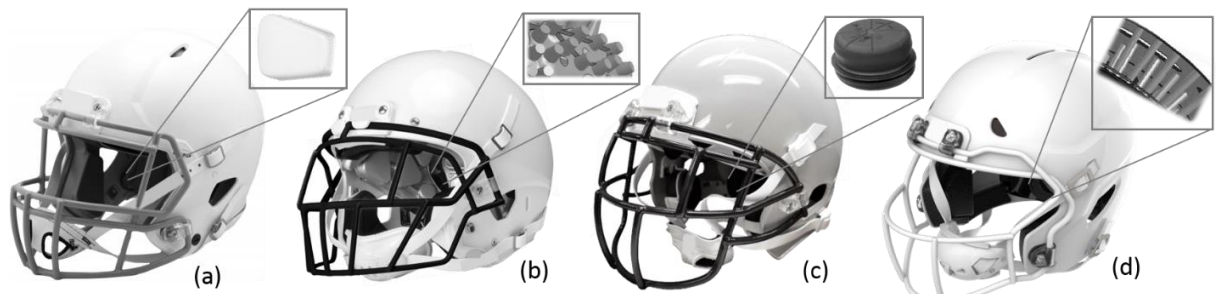

Fig. 1. Energy absorber technology of top performing commercially available helmets: a) foam, b) buckling cone, c) air damper, d) buckling beam.

In American football, all players at all levels need to use helmets which are certified by National Operating Committee on Standards for Athletic Equipment (NOCSAE) [28]. The current NOCSAE standard is based on drop tests and linear impactor tests whose measurements correspond to Gadd Severity Index (GSI) and peak linear acceleration, respectively. The majority of studies over the past few decades indicate that the main cause of concussion is inertial loading on the brain, which is mainly caused by head acceleration

[17], [29], [30]. During an impact, the head experiences both translational and rotational accelerations which, respectively, cause normal (i.e., pressure) and shear stresses in the brain tissue [31], [32], [33]. Due to the incompressibility and low shear modulus of the brain, the brain tissue is more prone to shear forces which are mainly the result of specifically rotational acceleration [34], [35] [36], [37], [38], [39]. In addition to the mandatory NOCSAE standard, there are also Virginia Tech (VT) safety ratings and NFL helmet rankings [40]. The VT helmet rating proposes a STAR value between 1 to 5 based on a Combined Probability of Concussion using rigid pendulum impact tests [41]. In contrast, the NFL helmet ranking recommends a Combined Metric (CM) based on rotational and translational kinematics of the head obtained through linear impactor tests [42].

In order to understand how the design of a football helmet plays a role in these aforementioned metrics and ratings, it is first important to understand the components associated with a football helmet. A football helmet typically consists of a shell, facemask, chin strap, chin cup, comfort pads, and energy absorber liner. In general, an energy absorbing medium can be made of either solid, gas, or liquid. Solid energy absorbers, such as closed and open-cell foams, demonstrate a deformation-dependent behavior where the damping force is mainly a function of the level of material deformation [43]. Gas energy absorbers, such as airbags, have a similar deformation-based response due to the compressibility of gas particles. Liquid energy absorbers have a unique property in that they exhibit a velocity-dependent behavior where the damping force is a function of the fluid velocity due to the incompressibility of the fluid [44]. Energy absorber systems in existing football helmets are typically designed using solid or gas, as shown in Fig. 1 [45], [46]: foam, air compression shock, and buckling beams or cones. In order to protect the head during a head impact, most of the energy needs to be absorbed by these energy absorbers in the helmet. However, the top 27 helmets in the NFL test standard show statistically significant differences in performance from the three top ranked helmets [25] [47]. This suggests that the performance of many available helmet technologies is still far below the maximum achievable performance, and that an entirely new approach may be necessary for significantly improving performance on these test beds.

Energy absorbers are responsible for protecting the brain by reducing the head acceleration, which is proportional to the force applied to the head through helmet. Hence, concussion risk can in principle be lowered by reducing the force transferred to the head. Considering that the peak head acceleration is one of the most widely used predictors of concussion, it is intuitively reasonable that for a given impact energy, a constant force over the full pad stroke can absorb the impact energy at the lowest force level, resulting in a minimum peak acceleration and thus minimal chance of injury. Therefore, to have a helmet with an ideal shock absorber, a constant force needs to be transferred to the head during impact [48]–[52]. A comparison between the force-displacement behavior of foam versus an ideal shock absorber is shown in Fig. 2.

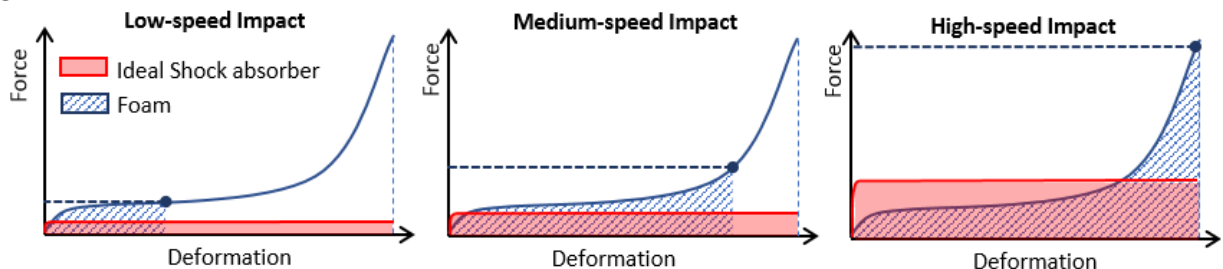

Fig. 2. (a) Force-displacement response of a foam versus an ideal shock absorber. The ideal shock absorber needs three features: (1) Force immediately rises to the required level, (2) Force level remains constant during compression, (3) Deform to maximize stroke length.

The criteria to predict the risk of brain injury can be categorized as: (1) kinematic criteria concerning solely motions of the head; or (2) stress/strain criteria based on brain tissue level loading [1]. The head kinematic criteria are proposed based on rotational and/or translational kinematics using either 3 degree-of-freedom (DOF) or 6 DOF measurements, e.g., Head Injury Criterion (HIC) [53], [54], Peak translational acceleration magnitude [41], Severity Index (SI) [55], Peak rotational acceleration magnitude [56], Brain Injury Criterion (BrIC) [57], and the brain angle metric (BAM) [58], [59], [60]. Using FE criteria, the tissue-level loading can be obtained and used in conjunction with strain-based metrics. The brain tissue strain and strain-based parameters are known as one of the primary concussion mechanisms. Some of the commonly used strain-based brain metrics are: maximum principal strain (MPS), [54], [61] and cumulative strain damage measure (CSDM) [62]. These FE criteria are used in conjunction with detailed FE models of the head.

This research work investigates the performance of football helmets in terms concussion prevention by considering the transmitted force to the head during an impact and some of the above-mentioned head kinematic criteria (HIC, peak angular acceleration, peak angular velocity) as well as FE criteria (MPS). First, the optimal damping force to absorb the impact energy from a ballistic mass is determined to show that the constant force profile does in fact minimize the brain tissue strain. Second, we propose a novel energy absorption technology of liquid shock absorbers to approach this minimum force level. Using finite element (FE) analysis, this optimal energy absorber is then integrated into a helmet. The NFL standard linear impact test is then simulated, and this new helmet's performance is compared with four other helmets with different energy absorption technologies (see Fig. 1) using kinematic metrics, brain FE criteria, and injury risk curves.

**Materials and Methods**

*Optimal force profile for football helmets*

We utilized a previously-developed reduced order model of the brain to find the optimal force profile that minimizes the strain in the brain. The Translational Head Injury Model (THIM) [63], [64] is a model used to study the effect of the impact force on the brain strain. This model is constructed using a Standard Linear Solid (SLS) model, representing the brain tissue, between two masses representing the skull and the brain, as show in Fig. 3.a. We will use this model to ultimately find the optimal force profile necessary to minimize brain tissue strain when absorbing the input impact energy. For this model, the state-space representation of the equations of motion (EOM) under force excitation is

$$\dot{x} = Ax + Bu, \quad x = [x_s \quad x_b \quad x_1 \quad \dot{x}_s \quad \dot{x}_s]^T, \quad u = F$$

$$A = \begin{bmatrix} 0 & 0 & 0 & 1 & 0 \\ 0 & 0 & 0 & 0 & 1 \\ k_1/c_1 & 0 & k_1/c_1 & 0 & 1 \\ -k_1/M_s & 0 & k_1/M_s & -c_2/M_s & c_2/M_s \\ k_1/M_b & 0 & -k_1/M_b & c_2/M_b & -c_2/M_b \end{bmatrix}, \quad B = \begin{bmatrix} 0 \\ 0 \\ 0 \\ 1/M_s \\ 0 \end{bmatrix} \quad (1)$$

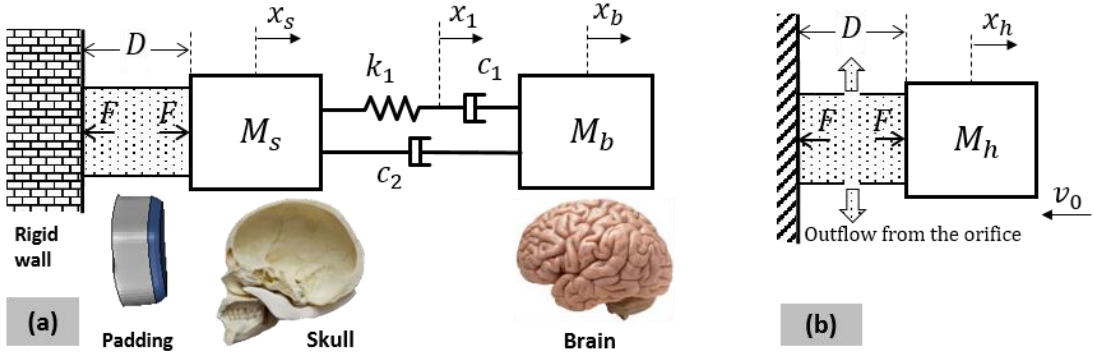

Fig. 3. (a) Dynamical model of the brain, skull and energy absorber. The energy absorber (dotted region) is located between the wall and the skull. (b) A schematic of a liquid shock absorber with a variable orifice size used to absorb the head impact energy. Over the entire impact duration $\ddot{x}_h > 0$, $\dot{x}_h < 0$, $x_h < 0$, where $x_h$ represents the head displacement as well as the deformation of the fluid shock absorber.

where $M_S$ is the mass of the brain, $M_B$ is the mass of the skull, $x_s, x_b$ are the skull and brain displacements, and $k_1$ and $c_1, c_2$ are the stiffness and damping coefficients of the skull-brain system, as shown in Fig. 3.a, respectively. The force transferred to the skull through the helmet, during an impact, is denoted by $F$. The state vector $x$ contains displacements and velocities of the brain and skull. The initial state is $x_0 = [0 \ \ 0 \ \ 0 \ \ -v_0 \ \ -v_0]^T$ where $v_0$ is the velocity of the helmeted head before hitting a rigid wall. The THIM model is one directional, and the model parameters depend on the impact direction. These parameters are given in [65], [63] for two common impact directions in football [63], i.e., front impact (A-P, Anterior to Posterior) and side impact (L–R, Left to Right). The magnitude of the relative brain-skull displacement $\delta(t) = |x_b(t) - x_s(t)|$ is assumed to be analogous to the brain strain. Consequently, the maximum of the relative displacement $max(\delta(t))$ is analogous to the brain Maximum Principal Strain (MPS) and is used as a metric for brain injury [59], [66], [67], [68].

An optimization problem is defined to find an optimal force which minimizes the relative brain-skull displacement as the helmeted head comes to rest in contact with ground. This type of impact represents the impact condition that might occur as a helmeted head strikes the ground from a fall. The injury criteria used as a measure of helmet performance is represented by

$$J(u) = \max_t \delta(t), \quad t \in [0, t_f] \quad (2)$$

where $J$ is the peak relative brain-skull displacement and $t_f$ is the duration of the force applied to the skull. In order to address the MPS criteria, the optimization problem is defined as the minimization of the peak relative brain-skull displacement subject to a set of conditions to be achieved at each $x_0 \leq x(t) \leq x(t_f)$ where $x_0$ is the initial value of $x$ at $t = 0$, and $t_f$ is the final time which is not known and needs to be determined by solving the optimization problem. The constraints are:

1. Transferring the head velocity to zero: Considering the contact between the skull and the helmet liner, this condition is applied on the skull velocity.

$$\dot{x}_s(t_f) = 0 \quad (3)$$

2. Respecting the maximum allowable displacement between the head and the helmet: The permissible displacement $D$ is the distance between the skull and the helmet shell in the impact direction and is determined by the thickness of comfort pads, and energy absorbers.

$$0 \leq x_s(t) \leq D, \quad t \in [0, t_f] \tag{4}$$

The optimal helmet energy absorber is formulated into the following optimal control problem in which the decision variable is $u$ and duration of the control process is $t_f$.

$$\min_{u, t_f} J(u) \text{ such that } \begin{cases} \dot{x}_s(t_f) = 0 \\ 0 \leq x_s(t) \leq D \end{cases} \tag{5}$$

This optimal control problem is solved by conversion into a constrained nonlinear programming problem [69]. To that end, the time interval $[0, t_f]$ and all continuous signals are zero-order hold discretized over $N$ subintervals [70], [71]. The discretized problem is solved using nonlinear optimization toolbox in MATLAB® [72]. The solution provides the optimal force together with impact duration. It is obvious that the solution depends on the initial velocity of the head $v_0$ and permissible displacement $D$. For a given range of initial impact velocities and permissible displacements, the problem is solved, and the results are shown in Fig. 6. Initial impact velocities are chosen based on the same velocities used in the NFL linear impactor helmet test protocol [42]. The solution to the optimal control problem confirms that a constant force over the entire given permissible pad displacement minimizes the relative brain-skull motion, regardless of the initial impact speed. Please refer to Results section for more information.

## Modeling of adaptive fluid shock absorber unit

In practice, an energy absorber with an optimal (constant) force profile can be generated using different physical systems. A fluid shock absorber with a variable orifice size is a good example of a system which can generate a constant force over its stroke length, yet is scalable with the initial impact velocity (in the same way that our optimization solution showed that the constant force scales with initial velocity – see Fig. 6.a,b). Different practical designs of fluid shock absorbers exist which can be adapted to helmet application and provide constant force [73], [74], [75], [76], [77].

If we look at the energy absorbed by the optimal constant force shown Fig. 6, we see that it is much higher than the energy absorbed inside the brain by dampers $c_1$ and $c_2$. Thus, the energy absorbed inside the brain by the dampers can be considered negligible in comparison with the energy absorbed by a constant-force energy absorber. Hence in the design and development of an optimal energy absorber, the head is assumed to be a single $M_h$ equal to $M_s + M_b$ with initial velocity $v_0$ and permissible displacement $D$ (see Fig. 3. b). The constant force required to absorb the head kinetic energy causes a constant head acceleration $\ddot{x}_h = v_0^2/2D$. Consequently the head velocity is $\dot{x}_h = -v_0(1 + x_h/D)^{1/2}$ and the required constant force is $F_{opt} = M_h v_0^2/2D$. Moreover, the pressure differential created by the orifice generates the hydraulic force,

$$F = k_h \left(\frac{\dot{x}_h}{A_o}\right)^2 \tag{6}$$

where $F_h$ is the hydraulic force, $\dot{x}_h$ is the relative velocity between the two ends of the fluid shock absorber, $A_o$ is the orifice area, and $k_h$ is the hydraulic constant. Considering the head kinematics, the fluid damper generates a constant force if $k_h/A_o^2 = M_h/2(D + x)$. Different combinations of $k_h$ and $A_o$ satisfy this relation. A practical combination of these parameters is considered.

$$k_h = M_h/2, \quad A_o = (D + x_h)^{1/2} \tag{7}$$

In the above equation, it is important to note that both design parameters are independent from the initial impact velocity $v_0$. In other words, the constant force generated by a fluid damper with the above parameters scales with the initial impact speed and there is no need to know $v_0$ to design the fluid damper. Moreover, the incompressibility of the hydraulic fluid increases the response time (i.e., rise-time) of the damping force (which is shown to be optimal as seen in the fast rise-times of the optimal force curves of Fig. 6 a,b). Having a fast force response (shown in Fig. 8) which scales with the impact velocity is the key feature of this fluid shock absorber which makes it different from current helmet energy absorbers. The damping force of the current helmet energy absorbers depends mainly on deformation distance which results in a slow rise-time; on the other hand, the damping force of the fluid shock absorber is velocity-dependent. In practice, the changing area of the orifice with damper deformation $x_h$ can be implemented with the techniques discussed in [73], [76].

### *Experimental testing of fluid shock absorbers*

To see how far off current helmet padding is from the optimal constant force profile, three helmet padding technologies were extracted from the side of football helmets to compare against the constant force. Foam, buckling cones, and air dampers were removed from newly purchased football helmets corresponding to the FE helmet models. The amount of tested padding was chosen to be the area of the padding covering the side of the head. All pads were tested using an instrumented drop set-up. Each pad was affixed to a 5 kg brass mass (approximately the weight of the human head) [78], [49]. A tri-axis accelerometer (356A66, PCB Piezotronics, Depew, NY) was affixed to the top of the brass mass to record impact acceleration. The mass was dropped onto a force plate (430_00_LCEL Force Plate, Cadex, QC) to record impact force. High speed video was recorded of each impact at 5000 Hz (Vision Research Phantom Miro LC320). All kinematic and force data were collected at 10 kHz and low-pass filtered with a fourth-order Butterworth 300 Hz cutoff. All pads were dropped from three heights: 50 cm, 100 cm, and 150 cm. This corresponded to final impact speeds of approximately 3.2 m/s, 4.2 m/s, and 5/5 m/s, as verified through high speed video footage. Each impact was repeated three times. Pad displacement was obtained by integrating acceleration twice, while force was obtained through the force plate.

### *Simulating fluid shock absorbers in football helmets*

To evaluate whether fluid shock absorbers could be utilized to improve the performance of commercially available football helmets, FE models are used to simulate the NFL linear impactor helmet test (see Fig. 1). First, the open-source, experimentally validated FE models for the four comparison helmets (see Fig. 4) were obtained from Biomechanics Consulting and Research, LLC (Biocore). These experimentally validated models were developed by Biocore with support from the NFL in collaboration with university partners [79], [80], [46]. These helmet FE models are simulated using LS-Dyna explicit FE solver (LSTC, Livermore, CA) by considering the solver version and precision level determined by developers of each helmet model. The pre-and post-processing is mainly preformed in LS-PrePost. The FE model includes the helmet, hybrid III head-neck, and linear impactor as shown in Fig. 4.b. To evaluate the performance of the helmet with liquid shock absorber, the open source helmet with buckling beam technology is modified to replace the energy absorption system with fluid damper elements. The fluid shock absorber is modeled using a discrete beam element with Material Type 70 in LS-Dyna environment (*MAT_HYDRAULIC_GAS_DAMPER_DISCRETE_BEAM). The hydraulic properties of the materials are obtained from eq. (7). The gas compression force is set to zero in order to only include the hydraulic effect.

The main elements of the modified helmet with fluid dampers are shown in Fig. 5, which include the shell, facemask, chinstrap, chin cup, jaw and ear padding, and fluid constant-force dampers. The helmet includes 25 fluid dampers distributed around the head where 22 dampers are distributed on the sides (i.e., 11 on the right side and 11 on the left side) and 3 dampers are on the sagittal plane. The fluid damper consists of two solid endcap plates and a discrete beam element representing a fluid damper with variable orifice coefficient [81]. The hydraulic properties of the damper are defined based on eq. (7) to generate constant force. The permissible displacement of the damper is the distance between the head and the helmet shell and varies between 35 mm to 40 mm depending on location of the helmet. The bottom endcaps are in contact with the head skin. The top endcaps are in contact with the helmet shell and are constrained to move on the axis defined by the orientation of the hydraulic element (see Fig. 5.d). The mechanical properties of the damper fluid and endcaps are those of hydraulic oil and polycarbonate shell, respectively. The total mass of the helmet is 2.014 kg. The weights of the helmets with foam, buckling cone, air damper, and buckling beam are 2.052 kg, 1.728 kg, 1.889 kg, 2.134 kg, respectively.

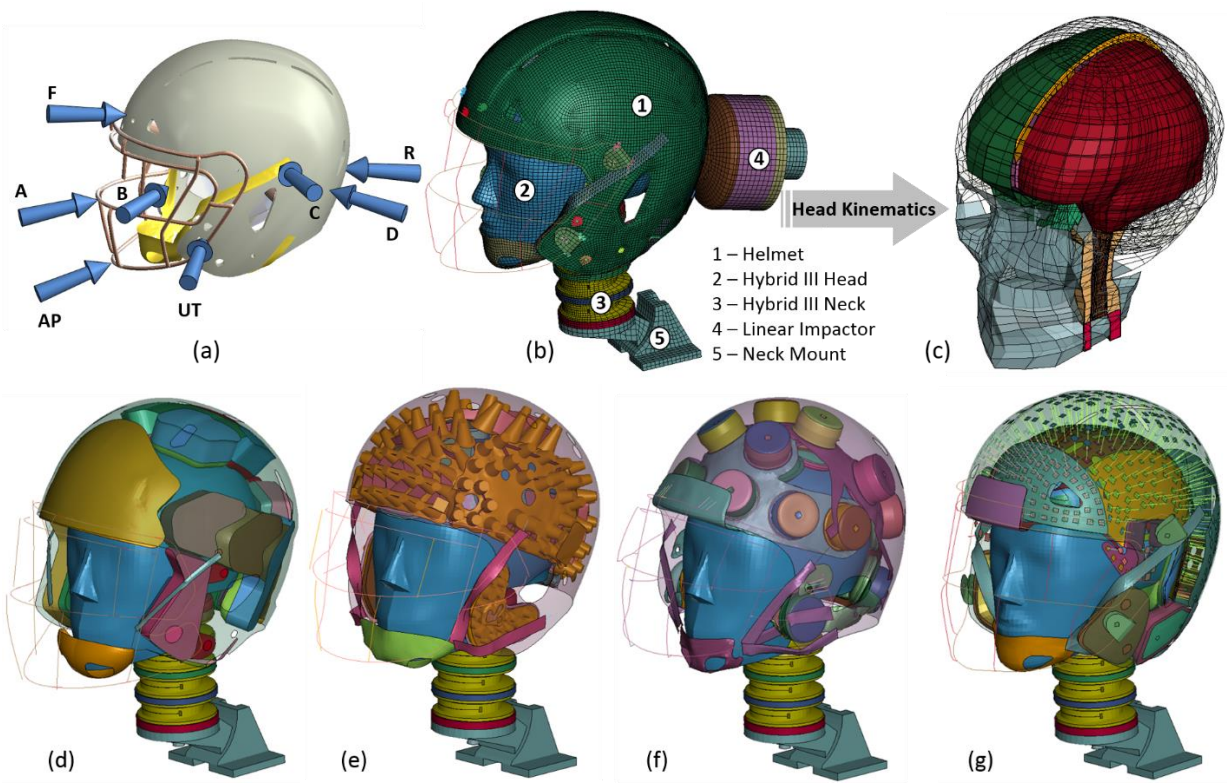

Fig. 4. Helmet FE modelling and simulation: (a) Location and direction of impacts given by NFL linear impactor helmet test protocol. At each impact position, the helmet is impacted at three different velocities. The head angle and the impact locations are provided in [42]. (b) Finite element model of the helmet with standard linear impactor [28], [79]. (c) KTH FE model of the brain and head. On the bottom, the FE models of four different helmets with different energy absorption technologies are shown [79]: (d) foam, (e) buckling cone, (f) air damper, and (g) buckling beam. The four FE models are available open-source [80].

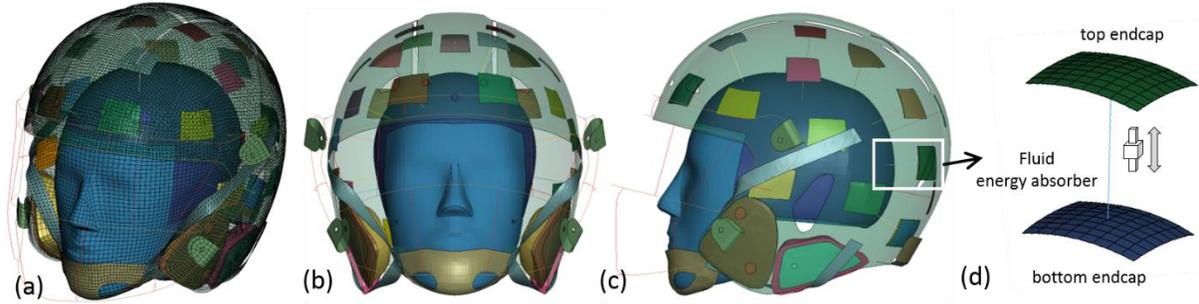

Fig. 5. FE model of the helmet with fluid shock absorber. The helmet shell is shown transparent to allow the shock absorbers to be viewed. (a) Helmet meshed FE model, (b) front view (middle), (c) side view, (d) Fluid energy absorber. The top endcap is in contact with the helmet shell and the bottom endcap is in contact with the head. Endcaps are constrained to be parallel and have linear motion along the damper element.

In the FE model, the linear impactor, hybrid III head-neck and all helmet parts except the fluid energy absorber are identical to those used in the model of the helmet with the buckling beam shown in Fig. 4.g and Fig. 1.d [79], [80]. The simulation outputs are: (1) kinematics of the head center of gravity (CG), (2) linear impactor ram acceleration, (3) linear impactor ram impact force. The head linear and rotational acceleration and velocity are used to evaluate the helmet performance. The standard NFL linear impactor helmet test protocol [42] includes 24 impact tests at 8 different locations, as shown in Fig. 4.a, and at three different impact velocities 5.5, 7.4, and 9.3 m/s, using a standard linear impactor system shown in Fig. 4.b. The brain FE model is developed by KTH Royal Institute of Technology in Stockholm, Sweden (hereafter referred to as the KTH model). The head FE model and the KTH brain model shown in Fig. 4.c is used to investigate the effect of helmet damper on the brain. The head kinematics obtained from the linear impact simulation is input to the KTH brain model in order to obtain the maximum first principle strain (MPS) [54]. The MPS is the ultimate output of the head simulation used to evaluate each helmet's performance [61].

*Statistical analyses to evaluate helmet performance*

The FE model of the modified helmet with fluid damper is simulated together with four other helmets with different energy absorption technologies. The results are obtained from simulations via LS-Dyna over 30 ms time intervals from the start of the impact [79]. For each helmet, the standard NFL helmet test and the KTH brain simulations are conducted for 8 different impact directions and 3 different impact velocities. Having 5 different helmets with different energy absorption technologies, 240 simulations are conducted in total. To evaluate the helmet performance in terms of prevention of brain injury, the 4 most widely used criteria are considered: MPS, peak rotational acceleration, peak rotational velocity, and HIC as previously discussed. In addition to different head injury criteria used to evaluate the helmet performance, the NFL introduces a helmet-specific performance metric called Combined Metric (CM) which is determined by conducting standard linear impactor test. Better performance of a helmet is represented by a lower CM. The CM is a kinematics-based metric obtained by combining rotational velocity, acceleration, and HIC [42], as shown in the equation below.

$$CM = \frac{1}{24}\sum_{i=1}^{24}\left(\frac{\alpha_{peak}(i)}{\alpha_{ave}} + \frac{\omega_{peak}(i)}{\omega_{ave}} + \frac{HIC15(i)}{HIC15_{ave}}\right) \qquad (8)$$

In this equation, $\alpha_{peak}$, $\omega_{peak}$, and $HIC15$ are peak rotational head acceleration, peak rotational head velocity, and HIC value over 15 ms time window, respectively. The grand average of all helmets over all test conditions for rotational acceleration, rotational velocity, and HIC are $\alpha_{ave}$, $\omega_{ave}$, and $HIC15_{ave}$, respectively. Using the linear impact simulation results, the CM of the helmets with different energy absorption technologies are compared in Fig. 12. The grand averages used for CM calculation are $\alpha_{ave} = 4127 \text{ rad/s}^2$, $\omega_{ave} = 38.0 \text{ rad/s}$, and $HIC15_{ave} = 208$ [42].

To approximate the clinical relevance of improving helmet performance using liquid shock absorbers, we estimated the risk of injury using six established concussion risk functions based on angular head impact kinematics. In previous work from our lab, we created injury predictors based on the peak angular acceleration vector, peak angular velocity vector, and the displacement of a 3 degree-of-freedom lumped parameter brain model [59]. Further, we used three other established injury risk functions based on peak angular acceleration magnitude [56], the Brain Injury Criteria [60], and the Virginia Tech Combined Probability metric [82].

For each simulated head impact of each helmet, we calculated the risk of injury as follows:

$$p_{injury_i} = \frac{1}{1 + e^{-\beta_o - x^T \beta}} \tag{9}$$

In this equation, $p_{injury}$ is the risk of injury at each location $i$, $\beta$ are the coefficients of the logistic regression fit, and $x$ is the injury criteria. The 95% confidence intervals for each injury prediction were calculated as follows:

$$\left[\frac{1}{1 + e^{x^T\beta - 1.96 \times SE(x^T\beta)}}, \frac{1}{1 + e^{x^T\beta + 1.96 \times SE(x^T\beta)}}\right] \tag{10}$$

$SE(x^T\beta)$ is the standard error of the computed probability, which is calculated using the covariance matrix $\Sigma$ of the logistic regression coefficients,

$$SE(x^T\beta) = x^T \Sigma x \tag{11}$$

Note, for both the Rowson and BrIC injury criteria, the covariance matrix was not reported and thus confidence intervals could not be calculated for each prediction.

For each helmet, to combine the risk of injury over all tested impact sites, we calculated the expected number of concussions if exposed to the 24 hits of the NFL test, as follows,

$$E_{concussion}(X) = \sum p_{injury_i} \tag{12}$$

Confidence intervals were propagated by calculating $E_{concussion}(X)$ using the upper and lower confidence intervals at each impact site.

## Results

The solution of the optimal control problem illustrated in Fig. 6.a and Fig. 6.b shows that for any given initial impact velocity and for any given permissible displacement, a constant force minimizes the relative brain-skull displacement. The duration of the impact $t_f$ which is part of the optimal control solution depends on the initial impact velocity ($v_0$) and permissible displacement ($D$). The dependence of the constant force level to $v_0$ and $D$ is shown in Fig. 6.c and Fig. 6.d. The oscillation of the skull velocity shown in Fig. 6.e is due to the low skull/brain mass ratio. The effective masses of the brain as skull for A-P impact direction are $M_S = 0.45$, $M_B = 4.09$ kg and for L-R impact direction are $M_S = 0.25$, $M_B = 4.09$ kg [65], [63]. Moreover, as shown in Fig. 6.e the maximum relative brain-skull displacement, which is the objective of the optimization problem, occurs at the end of the impact, i.e., $t = t_f$. Having residual brain velocity and displacement at the end of the impact ($t = t_f$) is not unexpected since the impact is assumed to be finished when the skull velocity reaches zero; nevertheless, the brain is lagging behind the skull and continues to oscillate until the residual energy is damped out. As far as the maximum relative brain-skull displacement $\delta(t)$ is concerned, for any time $t$ after the impact ($t > t_f$), due to damping behavior of the SLS brain-skull system and having no external force at $t > t_f$, the relative brain-skull displacement caused by the residual brain velocity, does not exceed the maximum relative brain-skull displacement generated during the impact, i.e., $\delta(t_f) > \delta(t > t_f)$.

To better understand the contribution of the optimal force in absorbing the head initial kinetic energy, the 6 different impact scenarios shown in Fig. 6.a,b with 3 different impact speeds and 2 different permissible displacement are considered. The parameters of the brain model in eq. (1) are obtained from [50], [48]. The average percentage of the energy absorbed by the optimal force is 96.24% ± 1.90, the average percentage of the energy absorbed by the internal brain dampers $c_1$ and $c_2$ is 3.42% ± 1.67, and the average percentage of the residual brain energy is 0.34% ± 0.23. The energy absorbed by the optimal force is much higher than the energy absorbed inside the brain by dampers $c_1$ and $c_2$.

To evaluate how far off existing helmet technology is from the ideal constant force, three different helmet pads were experimentally tested and the results are shown in Fig. 7. Among the three technologies, peak force was the highest for the air damper; however, all three were well above the constant force profile.

We evaluated how a liquid shock absorber, tuned for idealized performance in a benchtop setup, would perform when integrated into a football helmet. The operation of the fluid shock absorber inside the helmet can be investigated by looking at the damping force generated by an integrated fluid damper during an impact on the back of the helmet (i.e., R direction). The force-displacement behavior of a damper which is mainly involved in absorbing the energy is shown in Fig. 8. It has to be noted that the force levels shown in Fig. 8 should not be directly compared with Fig. 7 since the mass of the impactor and the test setups are different in these two studies.

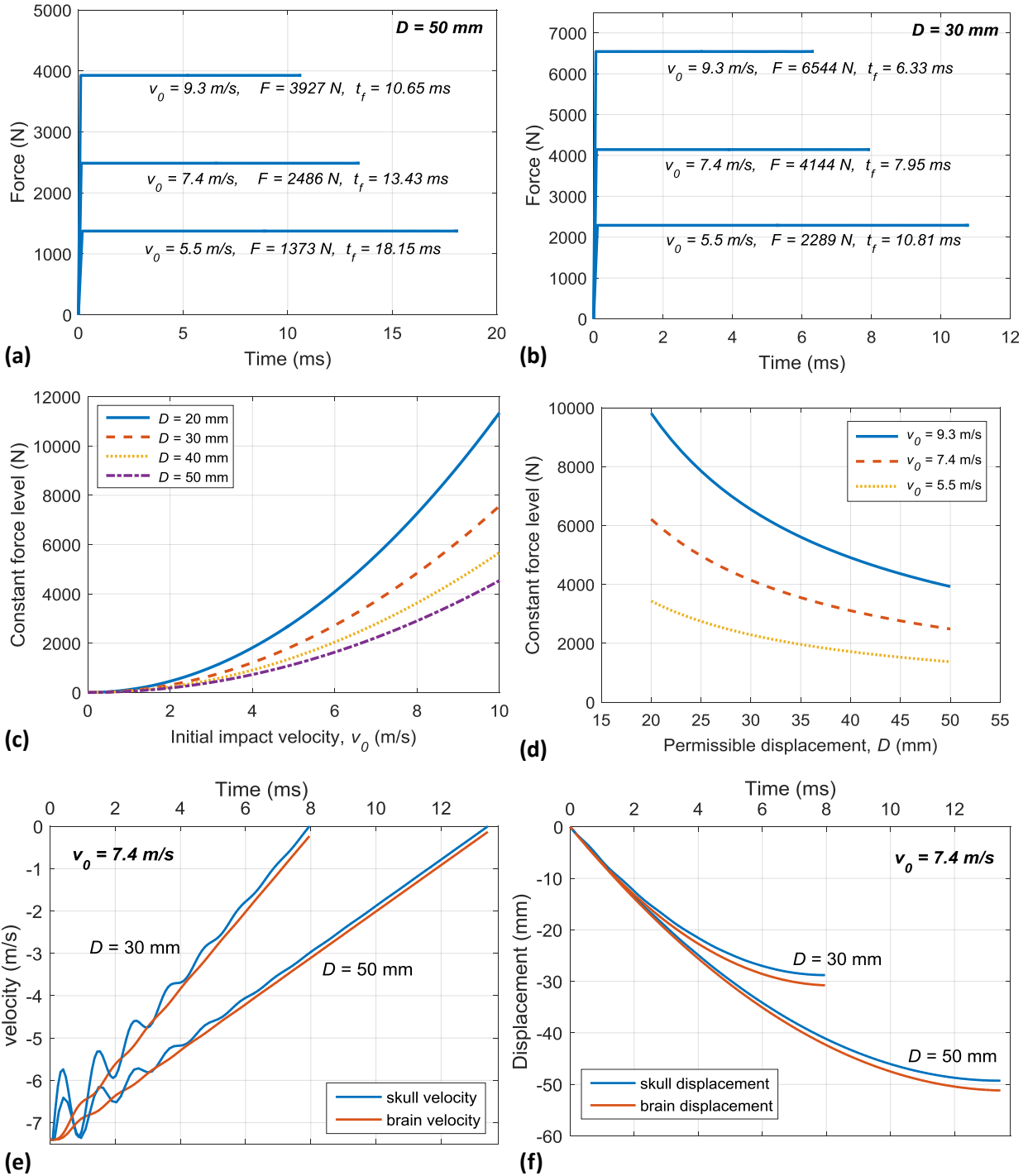

Fig. 6. Solution of the optimal control problem: (a) Optimal force for three different initial impact velocities with 50 mm permissible displacement, (b) Optimal force with 30 mm permissible displacement, (c) Optimal constant force dependency on initial impact velocity, (d) Optimal constant force dependency on permissible displacement. (e) Brain and skull velocity during the impact with an optimal force, (f) Relative brain-skull displacement during the impact with an optimal force. Due to the linearity of the brain-skull dynamical system, the relative displacements and velocities are only shown at a medium impact velocity and can be easily scaled for other impact velocities.

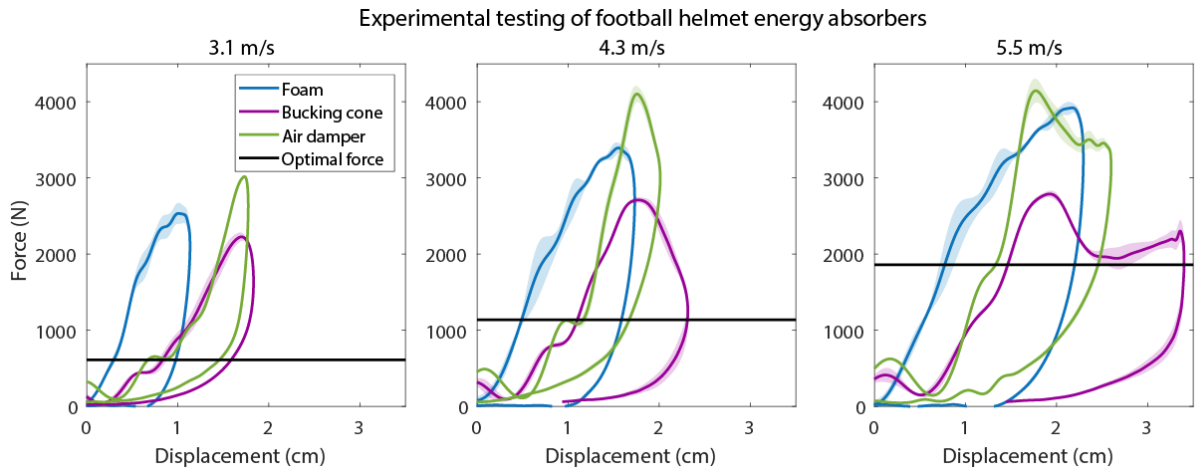

Fig. 7. Experimental testing of different helmet materials of approximately equal thickness. We removed foam, buckling cones, and air dampers out of the side of three football helmets. Each sample was tested under dynamic loading from a 5.0 kg weight. A fluid shock absorber is capable of maintaining an approximately constant force over all tested impact speeds, while existing energy absorbers provide force far above the optimal level. The solid line represents the mean over three trials, with the shaded region representing the standard deviation.

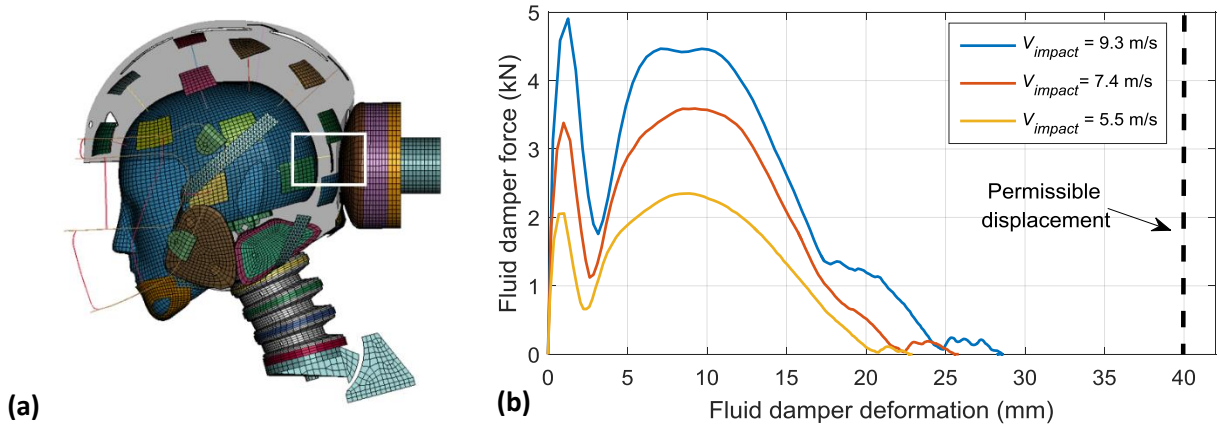

Fig. 8. (a) Rear impact configuration on the modified helmet with fluid shock absorber. (b) Damping force generated by a single fluid shock absorber located at the back of the helmet on the sagittal plane.

To better describe the procedure of helmet performance comparison in simulation, a sample of impact simulation and comparison in time-domain are demonstrated in Fig. 9, where the modified helmet with liquid shock absorber is compared with the foam helmet. For these two different helmets, the head kinematics obtained from the linear impact simulation is input to the KTH brain model in order to obtain the maximum first principle strain (MPS). The effect of liquid shock absorber on MPS variation and distribution over time is shown in Fig. 9 for a 9.3 m/s impact velocity and compared with the helmet with foam energy absorbers. The same simulation procedure is performed for all the 5 different helmet technologies discussed in this paper at 3 different impact velocities and 8 different impact locations. The summery of all simulations by considering different kinematic and FE metrics are shown in Fig. 10.

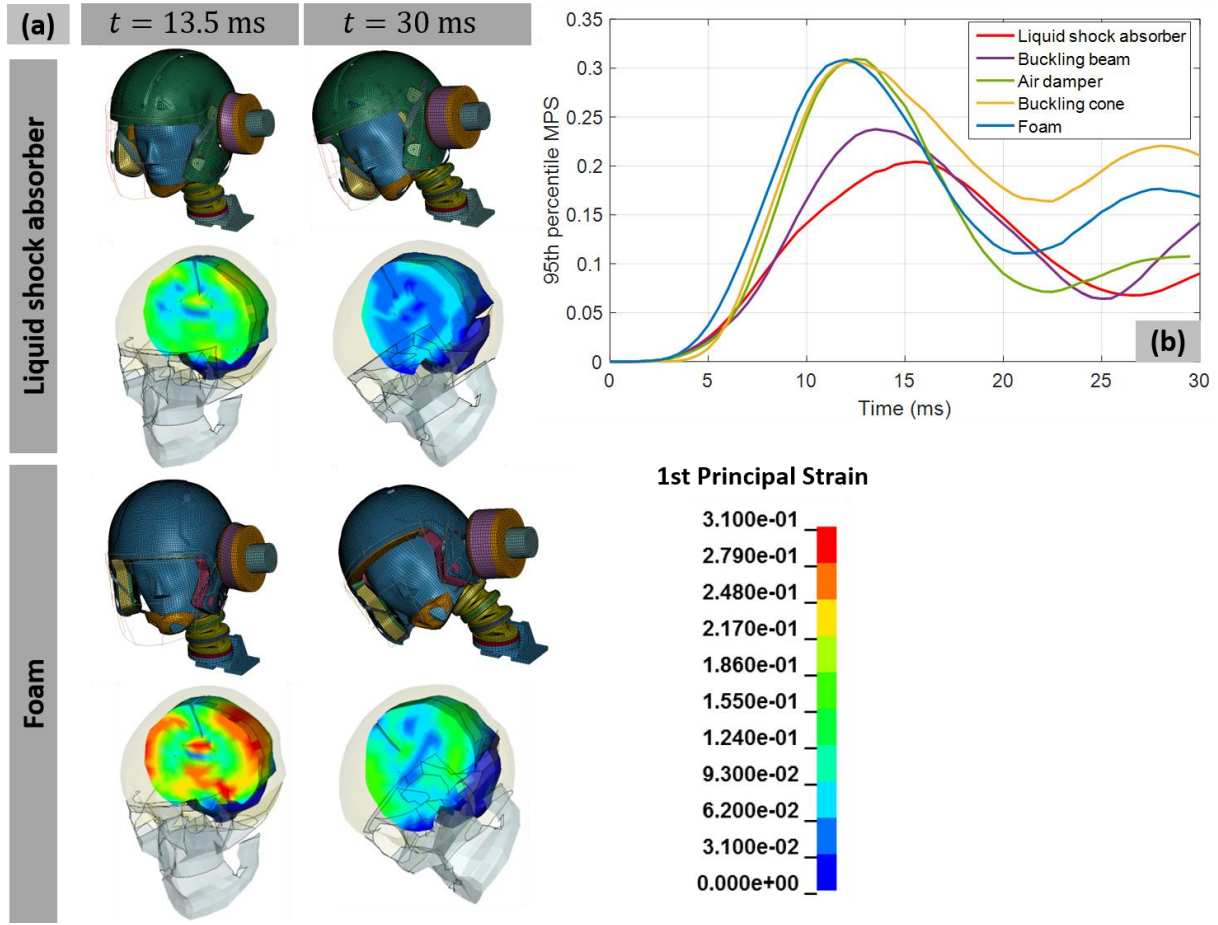

Fig. 9. Linear impact, lateral configuration at 9.3 m/s. (a) Distribution of the 1st principal strain inside the brain on mid- coronal section with transparent skull. The rest of the FE head elements are not shown to better illustrate the strain distribution. Comparison between the orientation of the head at $t = 0\ ms$ and $t = 13.5\ ms$ shows the MPS does not necessarily occur at large hear rotation. (b) MPS comparison over time between helmets with different energy absorption technologies. The strain distribution is shown for the two technologies with high and low MPS.

Results shown in Fig. 10 provide a comparison between the performance of the five different helmet FE models by considering a variety of different criteria at different impact speeds and locations. The fluid shock absorber helmet had the lowest peak rotational velocity, peak rotational acceleration, HIC15, and 95[th] percentile MPS at nearly all impact locations and speeds. To provide a general understanding about the overall helmet performance, the mean and standard deviation over different impact directions for each of these criteria at different impact speeds is shown in Fig. 11. We calculated the NFL combined metric for each simulated helmet and found that the fluid shock absorber helmet had the lowest value. Better performance of a helmet is represented by a lower CM.

Lastly, we computed the total number of expected concussions resulting from the 24 impacts of the NFL test for each of the helmets using four different injury risk functions (Fig. 13). For each of the tested risk functions, the fluid shock absorber helmet had a significantly lower predicted number of concussions than the existing helmets. Buckling beams showed the second lowest predicted number of concussions, while the foam helmet had the highest number of predicted concussions from the simulated NFL test.

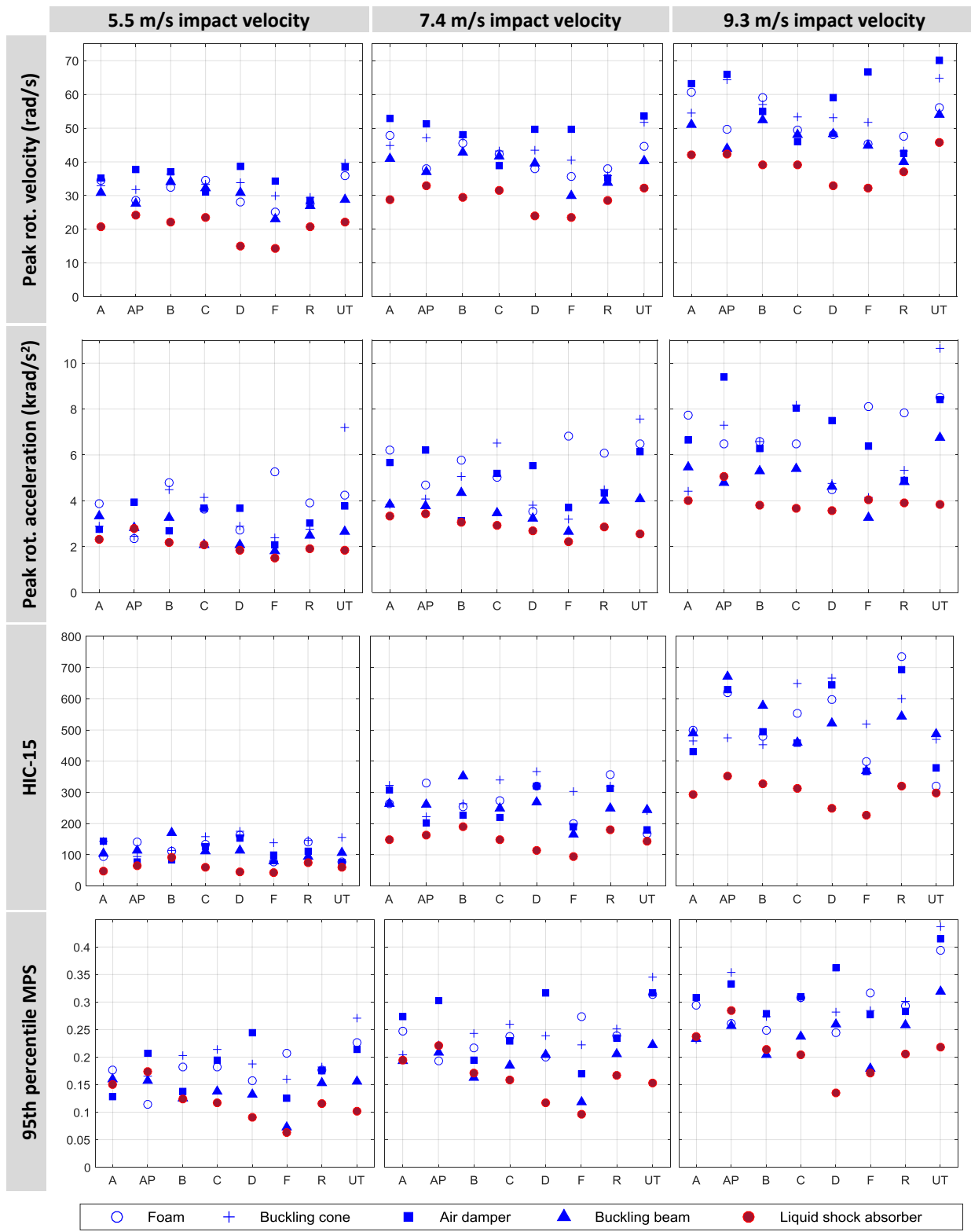

Fig. 10. Comparison between the performance of the helmets by considering different metrics: Peak rotational head velocity, Peak rotational acceleration, Head Injury Criteria HIC-15, and 95th percentile maximum principle strain, at 3 different impact speeds and 8 different impact locations.

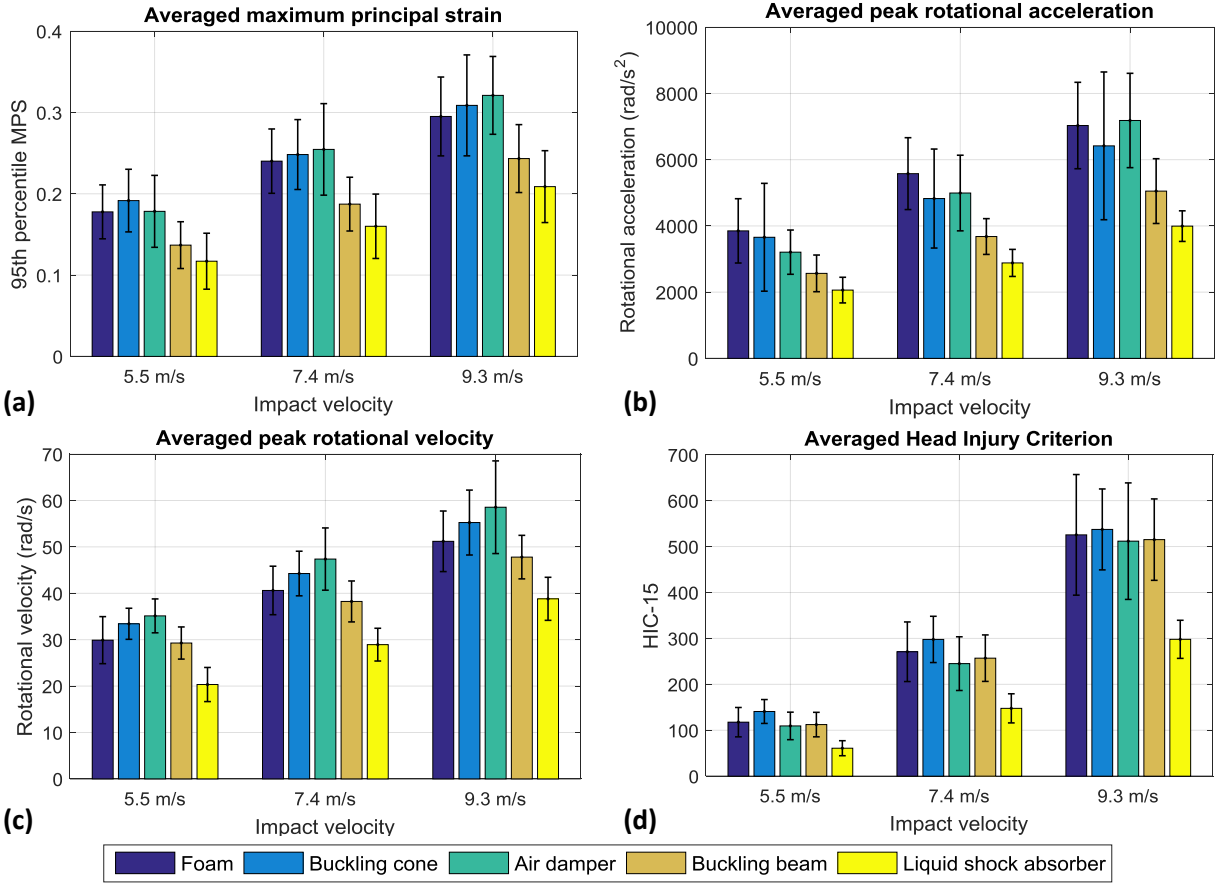

Fig. 11. Comparison between the performance of the helmets by the average of different metrics over impact directions

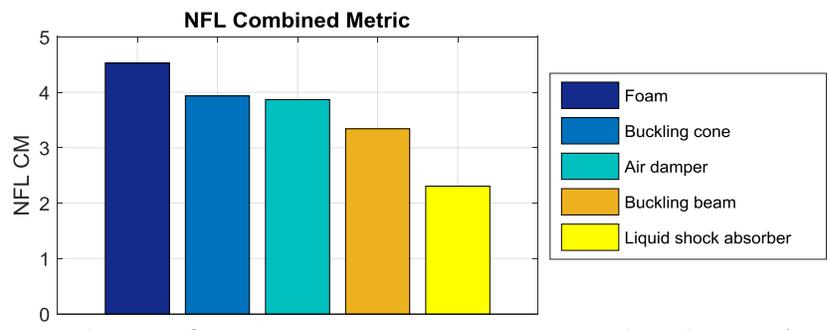

Fig. 12. Helmet performance comparison using NFL combined metric (CM)

## Discussion

This study shows that there is considerable room for performance improvement in football helmets to prevent mTBI. Considering MPS as one of the widely used brain injury predictors, it is shown that the MPS can be minimized by transmitting a relatively constant impact force to the head during the impact. The simulation results obtained from the both NFL helmet tests and the KTH brain model support our hypothesis that a liquid shock absorber which can provide a constant force during an impact can reduce the kinematic and FE metrics considered in this study: MPS, HIC, peak rotational acceleration and velocity. While other studies have suggested that constant force is the ideal force profile for helmets by reducing head acceleration [48]–[52], this study confirms that constant force also minimizes brain deformation according to both a rigid-body and finite element models.

The impact energy absorption in a helmet can be performed by three different mediums: solid (e.g., foam, buckling cone and beam), gas (e.g., air compression shock), and liquid. The damping force in solid and gas mediums is deformation-dependent which is a function of the level of material deformation; however, considering incompressible fluid mechanics, the damping force generated by a liquid is velocity-based which is a function of the fluid flow rate. A liquid shock absorber is introduced as a practical implementation of an energy absorber which is capable of generating constant force during an impact regardless of impact velocity. The key features of a fluid energy absorption system are the following. 1) Quick force response time proportional to the impact velocity, 2) Capability of maintaining constant force during the impact (Fig. 6), 3) Using the entire stroke length of the damper to absorb the energy. The force generated by the existing energy absorption technologies is deformation-dependent (i.e., deformation of a foam or a beam). This force-deformation behavior results in a low force level at the beginning of the impact due to the low amount of deformation (see Fig. 2). The force of these deformation-dependent materials rapidly increases as the pad compacts, causing very high forces towards the end of the impact, eventually reaching its deformation limit and "bottoming-out", at which point the helmet is no longer absorbing energy. Because of this, existing football helmets have to be designed for the worst-case scenario and are made to be extremely stiff to avoid bottoming-out in even the most severe impacts. However, this stiff padding means that force levels are much higher than necessary at lower speed impacts that can still cause injury. On the other hand, the force response of the liquid shock absorber is substantially more velocity-dependent, meaning it adapts its force displacement properties upon impact according to impact speed. Liquid also has the property of flowing out and, therefore, reduces the compaction problem with solid shock absorbers. Thus, the liquid shock absorber can react more softly in soft impacts, and more rigidly in severe impacts, while utilizing more of its stroke length than a traditional foam. Technologies based on gas shock absorber have been proposed or introduced for helmets, including the air dampers tested in this paper. While gas-based dampers are lightweight, the primary disadvantage of this approach is that gas is compressible and thus does not have the rapid response time of an incompressible fluid. The force created by these air dampers increases with the compression level, resembling spring behavior rather than dashpot behavior.

To evaluate the performance of an idealized fluid shock absorber in a helmet, we integrated fluid absorbers within a finite element model of a helmet and tested it against state-of-the-art football helmets in simulation. As expected, the fluid shock absorber provided reduced peak head kinematics in nearly every tested impact location and speed compared to the existing helmet designs. Further, among the eight standard impact directions shown in Fig. 4.a, four of them are on the helmet shell which are represented by R, C, D, F. In these impact directions the chinstrap is not in tension and no force is transferred to the facemask, so the facemask and the chinstrap have negligible effect on the energy absorption, and the impact energy is mainly absorbed by the dampers. Hence, a better way of evaluating the effect of energy absorption technology on the helmet performance is by excluding the impacts on the facemask (i.e., A, B, AP, UT). This can be done by only considering the R, C, D, F impact directions in the results shown in Fig. 10. In doing so, the average reduction in brain tissue strain will be 35.7% ± 8.3, which shows even better performance when compared with 27.6% ± 9.3 reduction which is obtained by considering all impact directions including impacts on the facemask. Hence, beyond helmet energy absorbers, there may also be room for helmet improvements on other aspects of the helmet such as the chinstrap, chin cup, or facemask. For example, new technologies such as rate-activated tethers have been proposed to help improve these components [48], [83].

The force-displacement provided in Fig. 8 confirms the key features of the fluid shock absorber which are quick force response, scaling damping force with the impact speed, and generating a damping force which is independent of the deformation level of the energy absorber element. Although we have validated that all of the dampers in the helmet provide constant force in an ideal benchtop scenario, these results show that these dampers do not provide a perfect constant force when integrated within a helmet. The main reasons for not observing constant force in our integrated helmet are the rotation of the head during the impact, the complexity and inherent nonlinearity of the head-helmet FE model in comparison with the simplified dynamical model eq. (1), and also connection of the head to a sliding carriage which causes the head-helmet system to move forward during the impact. However, having no dominant peak in the force response of the fluid dampers confirms the proper behavior of the damper. Moreover, the constant-force behavior of all single fluid damper elements used in the helmet model are verified by simulation in the ideal scenario shown Fig. 3.b.

Using a number of different risk functions, we evaluated the total number of expected concussions resulting from the 24 impacts of the NFL test standard (see Fig. 13). The angular velocity and angular acceleration metrics, taken from Laksari et al, were two of the top performing metrics on a dataset of over 900 impacts that included 27 clinical concussion diagnoses [59]. The angular acceleration metric from Rowson et al was based on 300,977 sub-concussive and 57 concussive head impacts taken from football players instrumented with helmet sensors. The BrIC AIS2 metric was based on 4,501 recorded head kinematics, of which 63 resulted in brain injury. These risk functions are based on entirely different datasets, and thus predict different risk levels; even so, the liquid shock absorber helmet is consistently the lowest in its risk of brain injury and number of expected concussions. This reinforces the notion that a substantial number of concussions could be prevented if energy absorption material in football helmets was more effective.

Taking these key results together, we found that there was approximately a 30% reduction in peak kinematics and MPS using the liquid helmet, which corresponded to a 75% reduction in number of expected concussions from the NFL test. To understand this observation, we plotted the angular acceleration magnitude risk curve [56], as well as the peak angular acceleration magnitude from each impact for each helmet (Fig. 13B). In this risk curve, there is a sharp increase in injury risk around 6000 rad/s$^2$, suggesting that there is a narrow threshold of injury for the human brain. Interestingly, testing conventional football helmets under the high speed NFL impact conditions resulted in accelerations just above this injury threshold. This could help to explain the high incidence of concussion in football. Further, the proposed liquid shock absorber helmet reduces accelerations just below this sharp injury threshold, explaining how even relatively small percent changes in angular kinematics can have considerable clinical significance in reducing concussion. However, it must be noted that sampling variability and errors in injury-reporting and kinematic sensors could have affected the development of the utilized risk curves, which would significant the number of predicted concussions [84].

This dataset of head impact kinematics and brain injury strain provides an opportunity to better understand the correlation between head kinematics and brain deformation in the context of football. The correlation between the MPS and the three kinematic criteria ($\alpha_{peak}$, $\omega_{peak}$, and $HIC15$) used to evaluate helmet performance is shown in Fig. 14. The angular acceleration has the highest level of correlation with MPS followed by angular velocity and HIC. Looking at Fig. 14, the low level of correlation between HIC and MPS might be due to the low body-to-head weight ratio of the linear impactor helmet test. The HIC is calculated based on head linear acceleration. In this test, the Hybrid III head-neck is connected to a sliding carriage having a mass of 17.7 kg which is considerably lower than the human body mass. This low body-to-head weight ratio may result in larger linear acceleration and lead to a HIC value which is not correlated well with MPS.

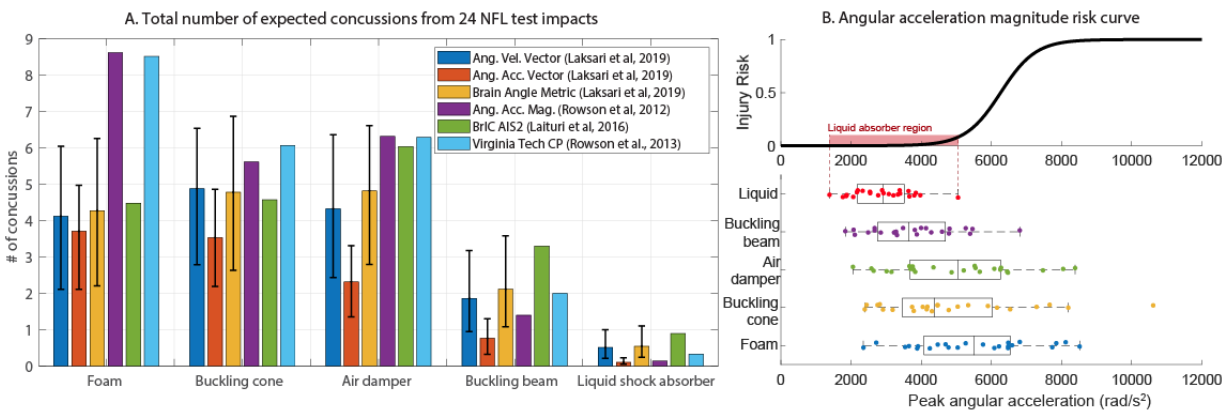

Fig. 13. Concussion risk for each of the helmets. A) Using four different concussion risk functions based on head impact kinematics, we calculated the total number of expected concussions resulting from the 24 impacts of the NFL test standard. The expected number of concussions for the liquid shock absorber was fewer than one for all risk functions, representing at least a ¾ reduction in concussions from the next best performing helmet. B) We plotted the angular acceleration magnitude risk curve, as well as the distribution of peak angular acceleration for each helmet. Liquid pushes the peak angular acceleration just below the bend of the risk curve, resulting in a dramatic reduction in predicted concussion risk.

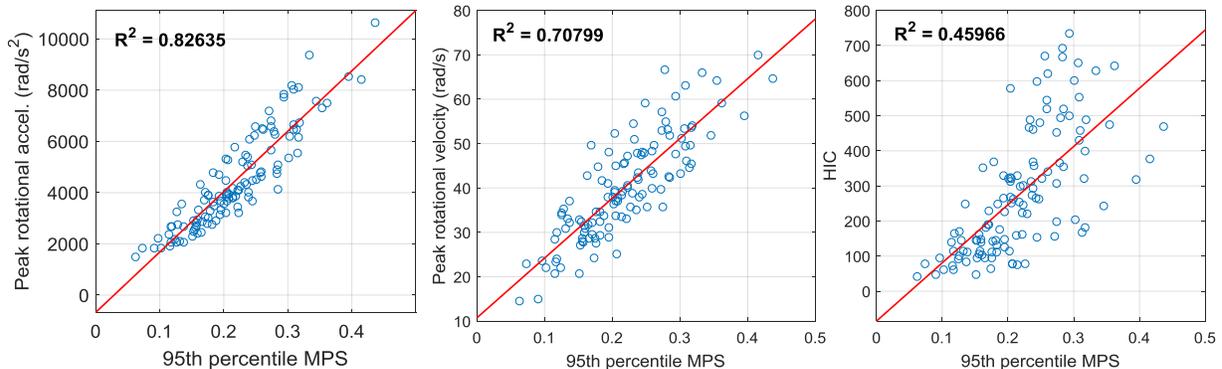

Fig. 14. Correlation between different kinematic criteria and the 95$^{th}$ percentile MPS which is one of the well-accepted TBI criteria. (left) Peak rotational acceleration (middle) Peak rotational velocity, (right) HIC.

There are a number of limitations to be noted. First and foremost, idealized constant force damping elements were used in the full helmet simulation, whereas some of the idealities may be non-feasible for a physical implementation. Further, as shown in the force-displacement plots, only 75% of the stroke length was used in a rear impact. This suggests that these dampers could be further tuned to use more of the stroke length and absorb the impact energy at a lower force level. By implementing shock absorbers within a physical helmet prototype, it is plausible that some of this performance could be sacrificed due to engineering constraints. For example, we considered each shock absorber to be a purely damping element, whereas a real fluid shock absorber prototype may have some elastic behavior due to the compression of the fluid vessel walls. Additionally, existing chinstraps and face masks were used in the modified fluid helmet, which may not represent the ideal performance. Future work will focus on investigating how these components could be modified to improve helmet performance. Future work will also focus on integrating physical fluid shock absorber prototypes into a helmet to experimentally validate these findings and also investigate the possibility of integrating fluid dampers into other parts of a helmet such as chinstrap, chin cup and even facemask. Regardless, it is valuable to know that modern football helmets remain well below the performance of the modified fluid helmet, and that fluid shock absorbers could prevent substantially more concussions if implemented in existing helmets.

A secondary limitation is that the helmet testing in this study was done exclusively on the NFL helmet test method conditions. These impact conditions may not accurately represent the initial conditions of real on-field impacts. For example, the top of the helmet is not evaluated in this test, although this region is likely important in protecting against cervical spine injuries [85]. Future work should focus on evaluating these helmets experimentally in other test scenarios. Lastly, the NFL test was run on computational FE models; these models have been extensively validated to correlate well with experimental time traces, but may have slight variations in peak kinematics [46], [79] [86]. However, we compare all results on the same platform, and thus we would expect to see the same trends if done experimentally.

*Acknowledgement*

We acknowledge the support of the Stanford Maternal and Child Health Research Institute. The study was partially supported by NSF Graduate Research Fellowship Program and the Ford Motor Company through the University Research Program (URP).